\documentclass[a4paper]{jpconf}
\usepackage{graphicx}
\begin{document}
\title{Intrusion Prevention and Detection in Grid Computing - The ALICE Case}

\author{Andres Gomez, Camilo Lara, Udo Kebschull for the ALICE Collaboration}

\address{Johann-Wolfgang-Goethe University, Frankfurt, Germany}

\ead{andres.gomez@cern.ch}

\begin{abstract}
Grids allow users flexible on-demand usage of computing resources through remote
communication networks. A remarkable example of a Grid in High Energy Physics (HEP) 
research is used in the ALICE experiment at European Organization for Nuclear Research 
CERN. Physicists can submit jobs used to process the huge amount of particle collision 
data produced by the Large Hadron Collider (LHC). Grids face complex security challenges. 
They are interesting targets for attackers seeking for huge computational
resources.
Since users can execute arbitrary code in the worker nodes on the Grid sites, special 
care should be put in this environment. Automatic tools to harden and monitor
this scenario are required. Currently, there is no integrated solution for such requirement. 
This paper describes a new security framework to allow execution of job
payloads in a sandboxed context. It also allows process behavior monitoring to detect intrusions, even when 
new attack methods or zero day vulnerabilities are exploited, by a Machine
Learning approach.
We plan to implement the proposed framework as a software prototype that will be tested as a 
component of the ALICE Grid middleware.
\end{abstract}

\section{Introduction}
ALICE (A Large Ion Collider Experiment) is a dedicated Pb-Pb detector designed to exploit 
the physics potential of nucleus-nucleus interactions at the Large Hadron Collider at CERN 
\cite{AliceProposal,AliceExperiment}. The ALICE experiment has developed the ALICE production 
environment (AliEn), which implements many components of the Grid computing technologies that 
are needed to analyze data. Through AliEn, the computer centers that participate in ALICE can 
be seen and used as a single entity - any available node executes jobs and file access is transparent 
to the user, wherever in the world a file might be\cite{GridTDR}.
\par 
Computing Grids allow the submission of user developed jobs composed by code and data. 
They interface with Internet and other communication networks, also with storage
systems and experiment infrastructure. They have challenging security requirements. Even when 
Grid system administrators perform a careful security assessment of sites, worker nodes, 
storage elements and central services, an attacker could still take advantage of unknown 
vulnerabilities (zero day). This attacker could enter and escalate her access privileges 
to misuse the computational resources for unauthorized or even criminal purposes. She could 
even manipulate the experimental data.
\par
In this document we focus on protecting and monitoring the job execution environment. 
Grids require tools to enforce the environment in such a way that user processes cannot 
access sensitive resources. They also require automatic tools to monitor job
behavior. These tools should analyze data generated in job runs like log entries, traces, system calls, 
to detect attacks on the system and react accordingly (for example sending alerts and 
stopping suspicious processes). This piece of software could be classified as
a Grid Intrusion Detection System (Grid-IDS). Traditional IDS perform
attack detection by fixed if-then rules based on signatures. This strategy
fails when innovative intrusion methods are used. We propose the use of Machine
Learning (ML) to overcome this inconvenience. Currently there is no framework that provides a 
solution for all above requirements. This documents provides a design for such
framework.
\par
This document is organized as follows: section~\ref{GridThreatModel} introduces
a Grid threat model. Section~\ref{IntrusionPrevention} explains a methodology to
provide security by isolation in the Grid. In section~\ref{IntrusionDetection} 
there is a detailed explanation of the Grid intrusion detection strategy we plan to apply. 
Section~\ref{Status} summarizes the project current state and the challenges
faced in the design and implementation of the desired methodology.
Finally, section~\ref{Conclussions} gives conclusions regarding the
proposal.

\section{Grid Threat model}
\label{GridThreatModel}
A threat model is useful to define a set of
security requirements and scope for a computing system. 
Threat modeling is a procedure for optimizing system
security by identifying objectives and vulnerabilities, and then defining
countermeasures to prevent, or mitigate the effects of threats to the system
\cite{ThreatModel}. The following subsections detail a threat model for Grid
computing.

\subsection{The goal of the adversary}
We call 'adversary' the individual that performs unauthorized actions 
to increase its assigned authorization and privileges, in order to misuse the
system resources or information \cite{adversary,Schneier}. 
In the Grid the adversary may have one or more goals, such as:

\begin{itemize}
  \item Steal sensitive data as private encryption keys, users certificates,
  tokens.
  \item Compromise users machines to distribute malware and steal valuable 
  user information.
  \item Carry out a Denial of Service attack.
  \item Abuse the Grid computational resources for criminal or not allowed
  activities, for instance to deploy botnets or mine crypto-coins.
  \item Damage the organization reputation by using resources to attack other
  organizations.
\end{itemize}

\subsection{Capabilities of the adversary}
To achieve the above mentioned goals, an attacker could take advantage
of several methods, such as:

\begin{itemize}
  \item Exploit software/hardware unknown or unfixed vulnerabilities.
  \item Listen to user network to gather sensitive clear text information.
  \item Perform a 'man in the middle' attack.
  \item Tamper other user jobs.
  \item Escalate his privileges.
  \item Access sensitive server configuration data.
\end{itemize}

\subsection{Specific Grid issues to be addressed}
\begin{figure}[h]
\begin{center}
\begin{minipage}{16pc}
\includegraphics[width=16pc]{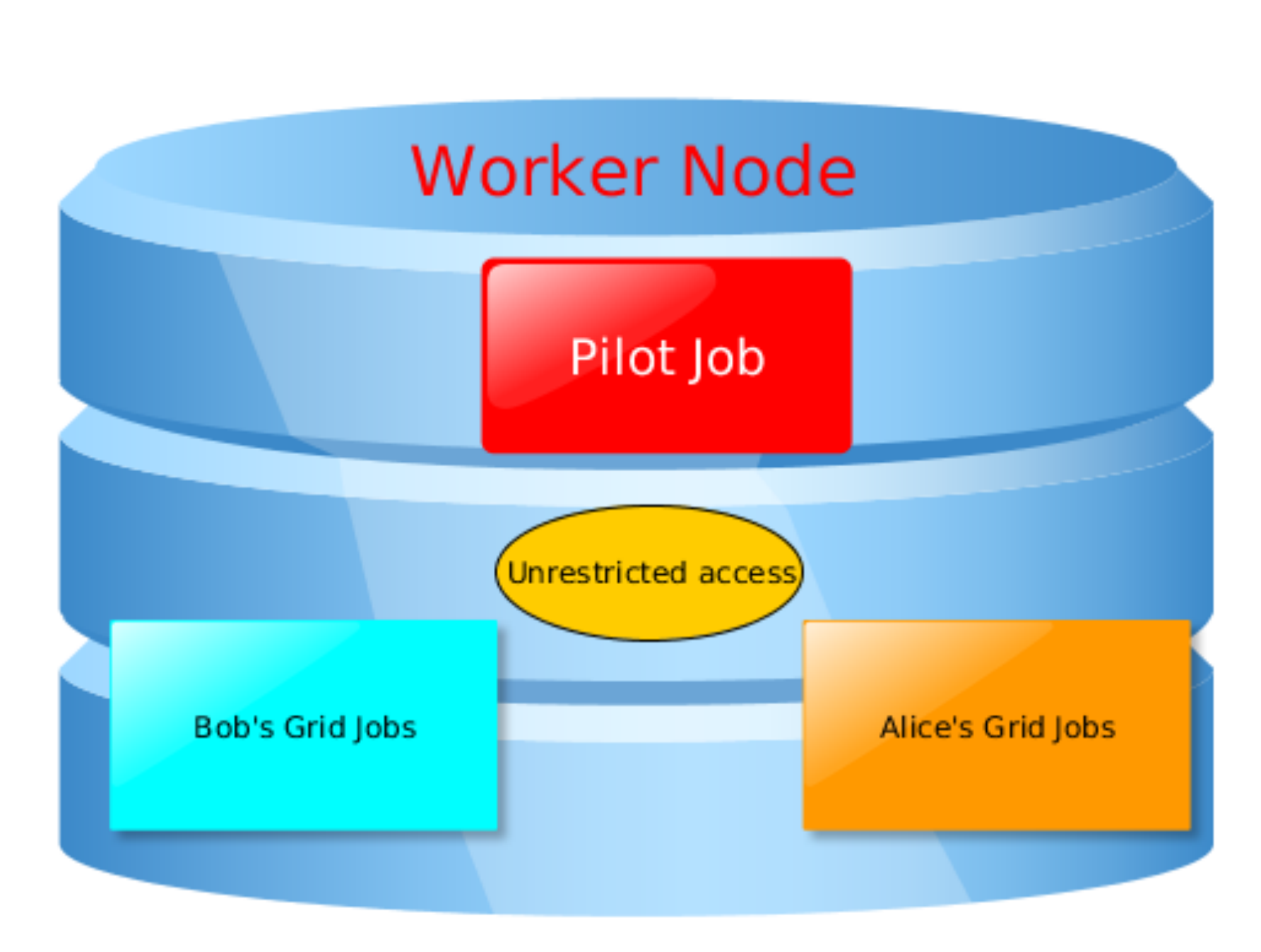}
\caption{\label{fig:figure1}Usual environment in Grid computing.}
\end{minipage}\hspace{2pc}%
\end{center}
\end{figure}

The security requirements in Grid computing are complex and challenging. 
We will focus in a set of issues related to the job execution environment 
inside the worker nodes. Commonly, jobs running in the worker nodes have far 
more access rights than required, sometimes restricted only by a machine local 
user account. Furthermore, if all the jobs are executed with the same user, it 
is easy for an attacker to tamper another user job. Usually these processes 
have access to sensitive server data (like /etc/passwd). Currently there is 
not an automatic method to prevent and detect intrusions when a job is 
attacking the Grid. Figure~\ref{fig:figure1} shows a schema of a common 
Grid execution environtment.

\section{Intrusion prevention}
\label{IntrusionPrevention}
We call intrusion prevention the hardening mechanism for the Grid execution environment. 
This mechanism guarantees that user processes cannot access privileged resources in the 
Grid site servers, such as server configuration files, other user jobs, the pilot job 
credentials and so forth. To accomplish this, we will use a security by
isolation approach.

\subsection{Security by isolation}
Security by isolation is a technique that enforces component separation 
(hardware or software) in such a way that if one of them is compromised, other
components remain safe \cite{isolation}. Several technologies provide
security by isolation, for instance Virtual Machines (VM), containers or even 
the Unix multiuser scheme. They provide various levels of isolation but the methods and 
constraints differ.
\par
\begin{figure}[h]
\begin{center}
\begin{minipage}{16pc}
\includegraphics[width=16pc]{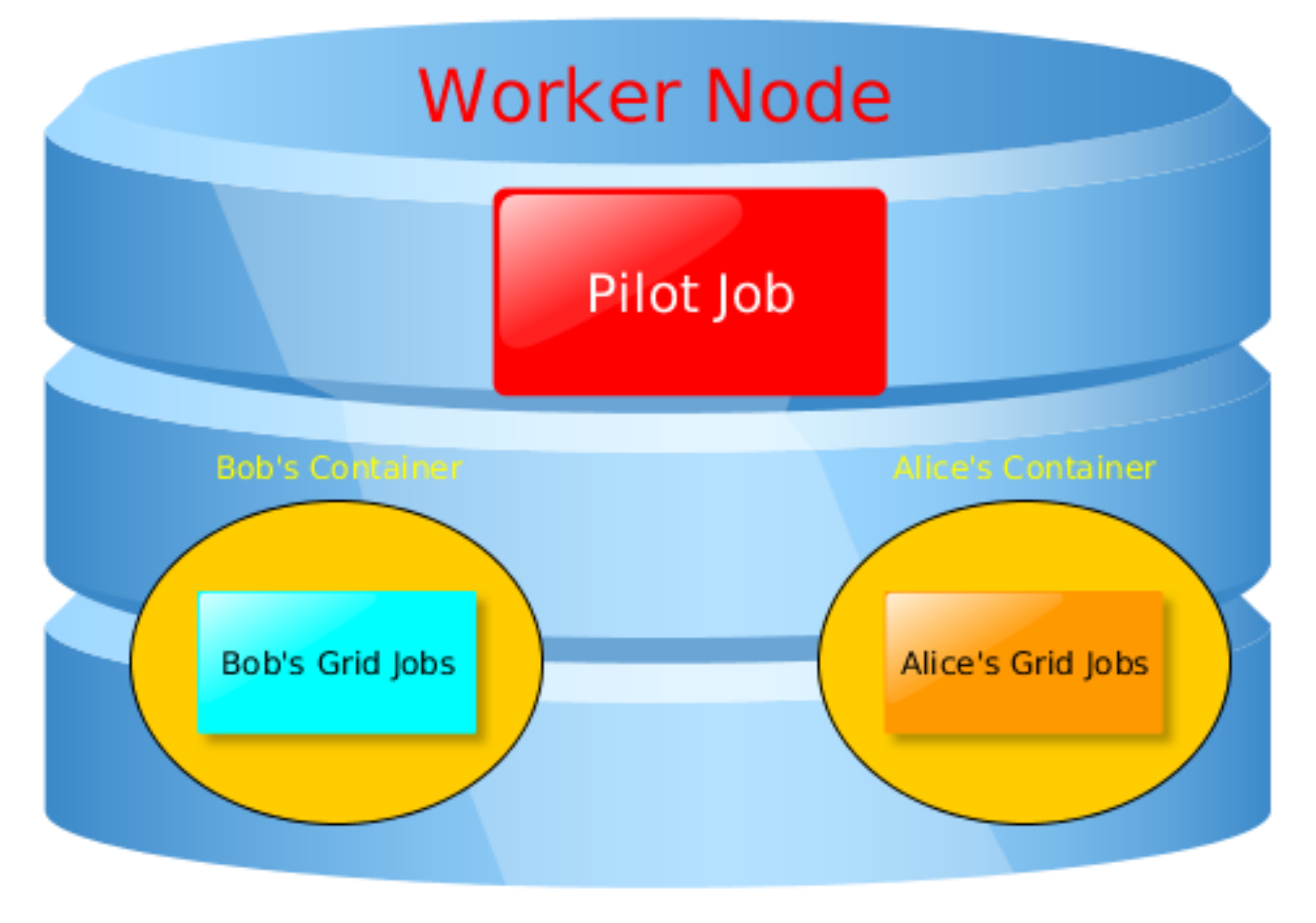}
\caption{\label{fig:figure2}Desired isolation scenario.}
\end{minipage}\hspace{2pc}%
\end{center}
\end{figure}

We pretend to enforce Grid site components so they run in a sandboxed
environment, with Grid unprivileged users for job execution. To achieve such goal, we propose 
the usage of containers. If needed, they still allow like-root access inside.
We could use tools like gLExec \cite{gLExec} as a complement. The next section
gives further details about this technology. Figure~\ref{fig:figure2} shows a scheme 
of the planned architecture of a worker node.

\subsection{Containers}
A container is a set of processes running on top of a common kernel (normally a Linux kernel) 
\cite{containers}. They are isolated from the rest of the machine and cannot affect the host 
or other containers. This technology uses namespaces to have a private view of the system 
(network interfaces, PID tree, mount points). It also uses Cgroups to have limited resource assignment. 
Containers can be seen as an extension of virtual memory space to a wide system scope.
\par
Containers provide a set of features that have advantages over other
virtualization technologies:

\begin{itemize}
  \item They are lightweight and fast.
  \item Virtual environments.
  \item Boot in milliseconds.
  \item Just a few MB of intrinsic disk/memory usage.
  \item Bare metal performance is possible.
\end{itemize}

\begin{figure}[h]
\begin{center}
\begin{minipage}{25pc}
\includegraphics[width=25pc]{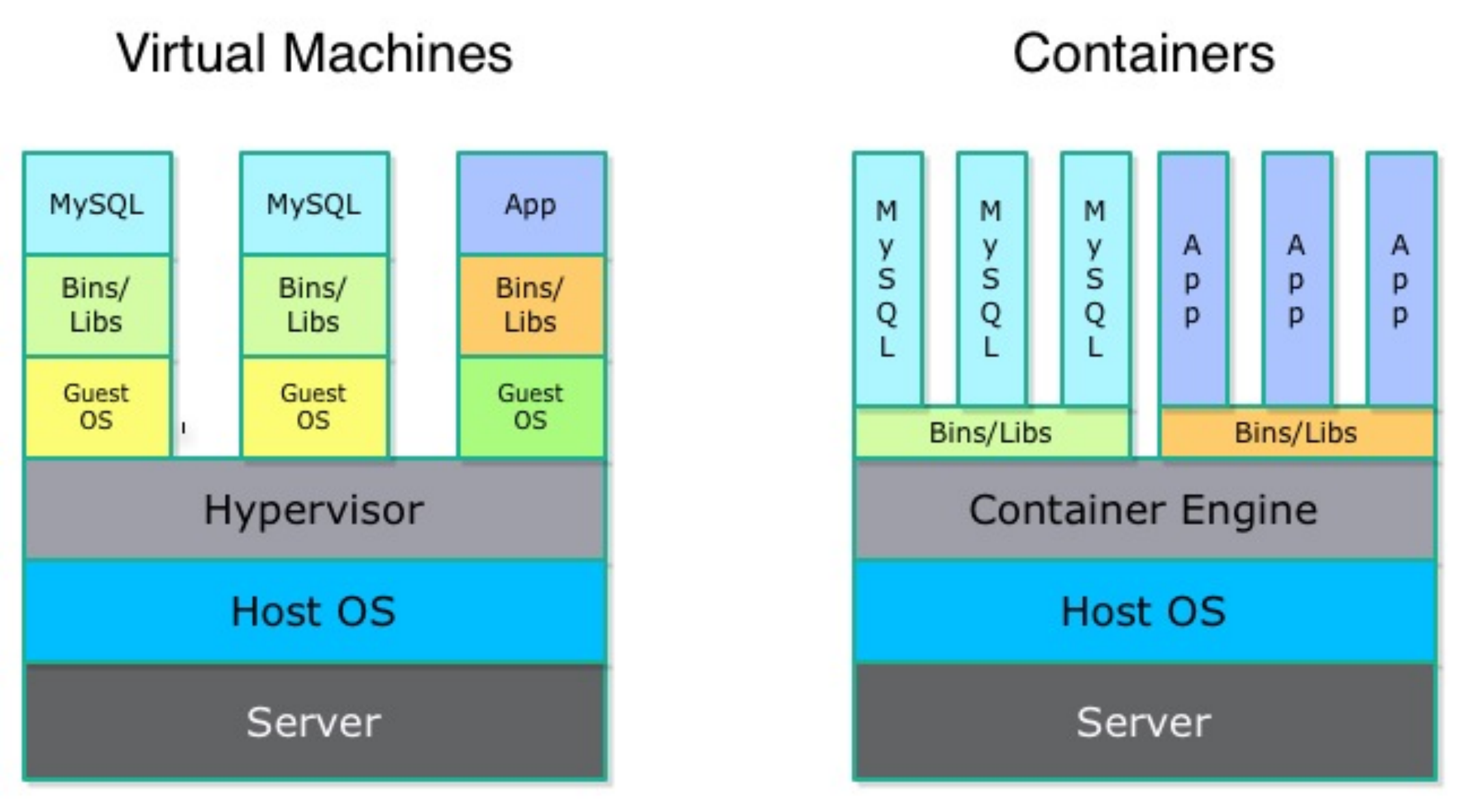}
\caption{\label{fig:figure3}Containers share a common kernel.}
\end{minipage}\hspace{2pc}%
\end{center}
\end{figure}

Figure~\ref{fig:figure3} shows how containers work together, isolated 
and sharing the same kernel. In opposition, VM have several 
kernels on top of the hypervisor.

When using containers for isolation, a security impact analysis is required. 
Commonly, VM are used in Grid and Cloud computing to achieve isolation. 
Then, a comparison of provided level of isolation between containers and Virtual
Machines is useful. Table~\ref{table:vmvscont} summarizes the main
security characteristics of both alternatives. It is not clear enough which alternative 
provides a higher level of isolation. However, container performance 
characteristics and its comparable level of security makes it a suitable 
selection for Grid computing. Moreover, it is possible to reduce the 
probability that a successful attack breaks the container isolation. 
This will be explained in the next section.

\begin{center}
\begin{table}[h]
\centering
\caption{Comparison of security provided on VM and Containers.}
\label{table:vmvscont}
\begin{tabular}{@{}l*{15}{l}}
\br
Virtual Machines&Containers\\
\mr
More isolation layers&The kernel is directly exposed\\
Big surface of attack&Less mature technology\\
Alone, it does not solve the Grid requirements&Reduced surface of attack\\
&Attenuation of kernel exposition possible\\
&Less time to update (kernel bugs)\\
&Fine-grained control\\
\br
\end{tabular}
\end{table}
\end{center}

\subsection{Reducing the surface of attack}
There are several alternatives to reduce the container surface of attack, 
decreasing the probability that a process escapes the enforced environment:

\begin{itemize}
  \item Use unprivileged user and containers \cite{unprivileged}.
  \item Use Seccomp-bpf to filter available system calls
  \cite{Seccomp-bpf}. This type of technology is currently being used in
  sanboxing methods for the Tor project, Firefox and Chrome browsers, and others.
  \item Use LSM technologies like Appamor \cite{apparmor}.
  \item Optionally use Grsecurity Linux kernel patch \cite{grsecurity}.
  \item Optionally we could still use containers over on top of Virtual
  Machines.
\end{itemize}

We plan to research on these topics to find out the best solution 
with the least performance overhead.

\section{Intrusion detection}
\label{IntrusionDetection}
\begin{figure}[h]
\begin{center}
\begin{minipage}{16pc}
\includegraphics[width=16pc]{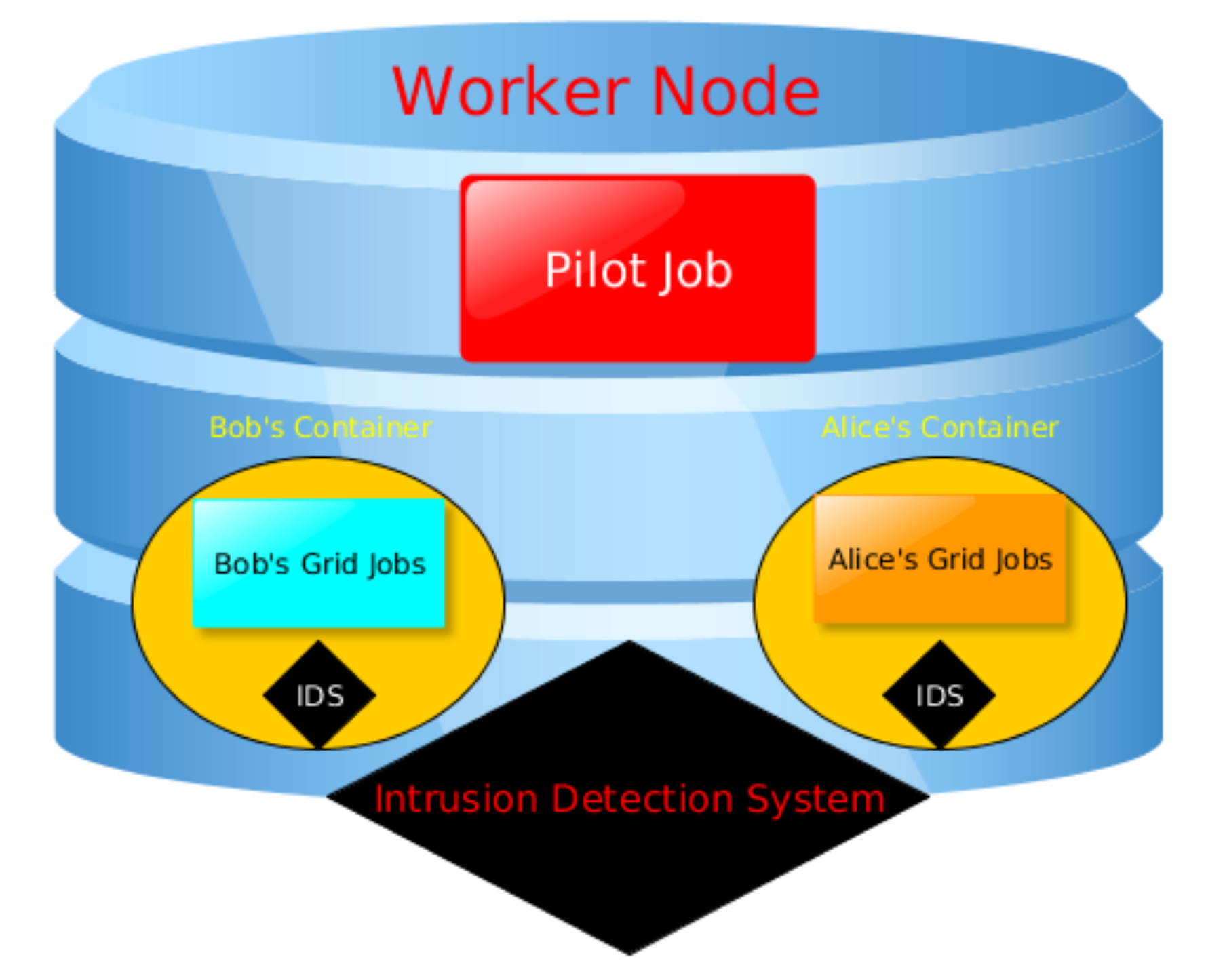}
\caption{\label{fig:figure4}Proposed Intrusion Detection in the worker nodes.}
\end{minipage}\hspace{2pc}%
\end{center}
\end{figure}

Even if we assumed that our execution environment is perfectly safe and 
it is impossible to scape the enforced isolation, there are many possible 
attacks that can affect the Grid. It is necessary yet to analyze the job 
behavior to determine when an attack on the system is occurring. If an user 
job is misbehaving,  the proposed framework should raise an alarm and perform 
predefined actions, for instance dropping the malicious processes. To detect 
innovative attacks we will use a Machine Learning algorithm. 
Figure~\ref{fig:figure4} shows the desired implementation of the proposed 
system regarding Intrusion Detection.

\subsection{Security metrics}
There are several alternative metrics for job behavior
analysis. Among them we can list the following: 

\begin{itemize}
  \item Job and system logs.
  \item System call sequence.
  \item Common monitoring data.
\end{itemize}

The goal is to select the best metrics that maximize relevant information about 
the job behavior without an evident decrease in performance. To analyze the
expected huge amount of information collected, intelligent algorithms are necessary. This allows 
to understand in a efficient fashion the security related events happening in the Grid.

\subsection{Machine Learning}
For job metric analysis we propose the usage of ML techniques. 
Common Intrusion Detection Systems use fixed rules and search for 
known attack signatures. However they have problems when unknown or 
slightly different intrusion methods are used. ML algorithms help to analyze big 
amount of data by learning the expected behavior and identifying abnormal situations. 
we plan to gather job metrics and classify them as 'normal' or 'malicious'. 
We will explore several alternative Machine Learning techniques, 
for instance Support Vector Machines (SVM) have been used successfully 
in security applications \cite{svm1, svm2}.

\section{Status}
\label{Status}
The project described in this document is currently in an early stage. 
At the moment we have deployed a full AliEn Grid installation in a single
server.
We have made several modifications to the framework to allow job execution 
inside unprivileged containers \cite{unprivileged}. There are several 
steps still to achieve. First we should create a custom site for security testing. 
Then a modification of AliEn/JAliEn to fully execute jobs in containers is necessary. 
We will research on ML methods for IDS and finally develop a complete 
prototype of the proposed framework.

\subsection{Challenges}
\label{Challenges}
There are challenges we might face in Grid security areas. For instance we should 
not try to increase the provided security if it means a big performance pitfall. 
In this paper we are considering only internal attacks, what about external
attackers? We also need innovative ways to analyze the huge amount of
trace/logs data generated in a efficient way. Maybe one of the most important challenges is to reduce the amount of 
false positives and false negatives, since the system administrators rely in the
accuracy of the security monitoring framework. Finally, for further research we need to consider 
what would be necessary in projects where Grid private data is required, for
example in the medical industry.

\section{Conclussions}
\label{Conclussions}
Computing Grids involve challenging security scenarios. Users can execute arbitrary
programs inside the organization infrastructure. Consequently this execution environment 
has to be hardened to limit the possibility of a successful intrusion. Containers provide 
security by isolation among the Grid components and the underlying system.
Besides, the automatic detection of attacks launched from the jobs is a priority task required to improve the overall 
security. To detect even innovative attack methods, a Machine Learning technique provides generalization 
ability to analyze job behavior information. We propose the design and development of security framework 
that fulfills such requirements. A software prototype will be developed and
tested as a component of ALICE Grid framework AliEn.

\ack
Authors acknowledge assistance from the ALICE offline team, 
specially Latchezar Betev and Costin Grigoras and the CERN 
security department specially Stefan Lueders and Romain Wartel. 
Authors also knowledge the German Federal Ministry of Education 
and Research (Bundesministerium fur Bildung und Forschung - BMBF) 
for its financial support.


\section*{References}
\begin{thebibliography}{9}
\bibitem{AliceProposal} 
The ALICE collaboration, ALICE - Technical Proposal for A Large Ion Collider Experiment at the CERN
LHC, CERN, Geneve, Rep. CERN-LHCC-95-71; LHCC-P-3, 1995.
\bibitem{AliceExperiment}
The ALICE collaboration, The ALICE Experiment at the CERN LHC, JINST, vol. 3,
no. 08, Aug. 2008.
\bibitem{GridTDR}
Eck, Christoph et al. LHC computing Grid: Technical Design Report. Version 1.06. 
20 Jun 2005. \textit{CERN, Geneva, Technical Design Report LCG}. 2005. 
Available from: https://cds.cern.ch/record/840543.
\bibitem{ThreatModel}
Wang L, Wong E, Xu D. A threat model driven approach for security testing. 
\textit{Proceedings of the Third International Workshop on Software Engineering
for Secure Systems} [Internet].
IEEE Computer Society; 2007 [cited 2015 May 6]. p. 10. 
Available from: http://dl.acm.org/citation.cfm?id=1269070
\bibitem{adversary}
Modi C, Patel D, Borisaniya B, Patel H, Patel A, Rajarajan M. A survey of
intrusion detection techniques in Cloud. \textit{Journal of Network and Computer
Applications}. 2013 Jan;36(1):42-57.
\bibitem{Schneier}
Schneier, Bruce. Applied Cryptography (2Nd Ed.): Protocols, Algorithms, and
Source Code in C. 1995. 0-471-11709-9. \textit{John Wiley \& Sons, Inc.}. New
York, NY, USA.
\bibitem{isolation}
Steve Mansfield-Devine. Security through isolation. \textit{Computer Fraud \& Security}. 
5, 8 - 11. 2010. 1361-3723. 
Available from: http://www.sciencedirect.com/science/article/pii/S136137231070052X.
\bibitem{gLExec}
Groep, David, Oscar Koeroo, and Gerben Venekamp. gLExec: gluing grid computing to the Unix world. 
\textit{Journal of Physics: Conference Series}. Vol. 119. No. 6. IOP Publishing,
2008.
\bibitem{containers}
Stephane Graber, Ubuntu Foundations Team [homepage on the Internet].
LXC 1.0: Blog post series; 2014 [updated 2014 Jan 17; cited 2015 May 11].
Available from: https://www.stgraber.org/2013/12/20/lxc-1-0-blog-post-series/
\bibitem{unprivileged}
Stephane Graber, Ubuntu Foundations Team [homepage on the Internet].
Introduction to unprivileged containers; 2014 [updated 2014 Jan 17; cited 2015 May 11].
Available from: https://www.stgraber.org/2014/01/17/lxc-1-0-unprivileged-containers/
\bibitem{Seccomp-bpf}
The Linux Kernel Archives [homepage on the Internet]. SECure COMPuting with
filters [cited 2015 May 11]. 
Available from: https://www.kernel.org/doc/Documentation/prctl/seccomp\_filter.txt
\bibitem{apparmor}
Ubuntu wiki [homepage on the Internet]. AppArmor [cited 2015 May 11]. 
Available from: https://wiki.ubuntu.com/AppArmor
\bibitem{grsecurity}
Grsecurity [homepage on the Internet]. Grsecurity [cited 2015 May 11]. 
Available from: https://grsecurity.net/
\bibitem{svm1}
Yuping L, Weidong L, Guoqiang W. An Intrusion Detection Approach Using SVM and
Multiple Kernel Method. \textit{International Journal of Advancements in
Computing Technology}. 2012 Jan 31;4(1):463-9.
\bibitem{svm2}
Gascon H, Yamaguchi F, Arp D, Rieck K. Structural detection of android malware
using embedded call graphs. \textit{ACM Press; 2013} [cited 2014 Nov 17]. p.
45-54.
Available from: http://dl.acm.org/citation.cfm?doid=2517312.2517315
\end{thebibliography}
\end{document}